\documentclass[12pt]{article}

\usepackage{epsfig}

\title{Energy Scavenging in Silicon Raman Amplifiers}
\author{S. Fathpour, K. K. Tsia, and B. Jalali\\
 {\it Electrical Engineering Department} \\{\it University of California, Los
 Angeles}
}

\date {}
\begin{document}
\maketitle

Continuous-wave Raman amplification in silicon waveguides with
negative electrical power dissipation is reported. It is shown that
a p-n junction can simultaneously achieve carrier sweep-out leading
to net continuous-wave gain, and electrical power generation. The
approach is also applicable to silicon Raman lasers and other third-order
nonlinear optical devices.

\pagebreak

Silicon-on-Insulator (SOI) has been long recognized as a preferred platform
for the realization of both electronic and photonic devices.$^1$ One of the
most
attractive features of the SOI material system is the prospect of full
integration of optical and electronic devices on the same substrate.
Much progress has been made in Si-based photonics towards low-loss waveguides,
photodetectors, electro-optic modulators, light sources and optical
amplifiers.$^2$ However, little or no attention has been made to the power
dissipation of photonic devices. At the same time, the problem of power
dissipation in silicon VLSI is so severe that it threatens to bring to
 halt the continued advance of the technology, as described by the
 Moore's law.$^3$ This fact is highlighted by the recent momentous shift of
 the microprocessor industry away from increasing the clock speed and in
 favor of multi-core processors.$^4$ Evidently, realization of low-power silicon photonic devices is essential for opto-electronic integration.

Typically, lasers are the most power-hungry photonic devices. However, the lack
of an electrically-pumped Si laser, to the date, dictates an architecture
where the light source remains off-chip. In such architecture, an off-chip
source empowers the chip, whereas modulators, amplifiers, photodetectors, and
perhaps wavelength converters, are integrated on the chip. Among these devices
the optical amplifier has the highest power dissipation. To the date, Raman
amplification has been the most successful approach for achieving both
amplification and lasing in silicon.$^{5-12}$ The main limitation of the silicon
Raman amplifier is the optical loss caused by scattering from free carriers
that are generated when the high intensity pump causes two photon absorption
(TPA).$^{10,13}$  To achieve net continuous-wave gain, a reverse biased p-n junction
can be used to sweep the carriers out of the waveguide core.$^{10-13}$ However,
this comes at the expense of significant electrical power dissipation.$^{14}$ The
present work addresses this power dissipation in silicon Raman amplifiers.
It is shown that at moderate gain levels, it is not only possible to avoid
the power dissipation, but also to extract net electrical power from the
device. The net electrical power generation is achieved by operating a p-n
junction diode, which straddles the waveguide, in the fourth quadrant of its
current-voltage (I-V) characteristics. In this mode, the TPA-generated
carriers are swept out by the built-in field of the junction, yet the device
delivers electrical power. The concept becomes clear if one considers the
device as a nonlinear optical equivalent of a solar cell.

A schematic of the device is shown in Fig. 1. Devices were fabricated
by standard optical lithography, dry etching, dopant implantation, passivation,
and metallization techniques. Different variations of waveguide widths
(W =1.5 to 3.0 $\mu$m) and spacing between doped region wells and ridge edges
(d =1.6 to 2.8 $\mu$m) were obtained.  Laterally-tapered mode-converters were
fabricated in order to decrease the coupling loss into and out of the
waveguides. The facets were polished but left uncoated. The best experimental
results, presented in the following, are for 3-cm long waveguides (including
the tapers) with W=1.5 $\mu$m and d=2 $\mu$m.  A linear propagation loss of ~0.5 dB/cm is measured in these waveguide using the Fabry-Perot technique.

The pump laser is a New Focus external-cavity tunable diode at 1539 nm with a
linewidth of 0.1-0.2 nm amplified with an EDFA. The Stokes signal is a 1673
nm DFB laser biased at an output power of ~4.5 mW. The pump and Stokes beams
are coupled in and out of the waveguide with two identical objective lenses
(M=20X, 0.40 NA), after combining the two beams in a wavelength division
multiplexing coupler. Also, a 5/95$\%$ tap coupler is incorporated into the pump
optical path to measure its power. The electrical loop that biases the diode
consists of a power supply, a 1 k$\Omega$ series resistor and a current
meter. Unlike previous reports by us and others,$^{11,12,15}$ no thermoelectric
cooler was required to stabilize the device temperature.

Figure 2 presents the measured on-off Raman gain at different coupled pump
powers and biasing conditions. Although TPA of the pump power is not a major
detriment by itself,$^{10}$ the generated free carriers increase the loss
substantially, preventing Raman gain in silicon waveguides.$^{13}$ This is in
agreement with the optical loss observed in Fig. 2 for an open-circuit p-i-n
junction. When a reverse bias is applied, the induced electric field removes
the TPA-induced free carriers from the waveguide region, hence reducing the
carrier effective lifetimes. A maximum on-off gain of ~4 dB is obtained at a
reverse bias of 15 V. Considering the measured linear loss of 0.5 dB/cm, the
highest achievable net Raman gain is thus ~2.5 dB in the 3-cm long device.
Higher reverse voltages do not increase the gain considerably, which may be
due to saturation of the drift velocity. This argument is supported by the
fact that for a p- and n-well separation of 5.3 $\mu$m in the device, the average
electric field at 15 V is equal to the peak electric field value of ~3$\times 10^4$ V/cm in silicon. It is also observed that an on-off Raman gain of 2.7 dB is
attained when the diode is short-circuit (0 V).  This corresponds to a net
Raman gain of ~1.2 dB, which is higher than our previous report at zero bias
where no lateral mode-converting tapers was involved.$^{15}$

As seen in Fig. 2, on-off Raman gains as high as 2 dB is measured when the
device is forward-biased at voltages $\le$0.7 V. The importance of this biasing
regime is that power dissipation is negative. Figure 3 shows the generated
power for biases of 0.6 and 0.7 V at different pump intensities, extracted
from the measured current and voltage drop across the diode. The TPA-induced
photovoltaic effect is also clearly evident in the measured I-V
characteristics presented in Fig. 4. The attenuated pump intensity via TPA
creates free carriers in the SOI waveguide. The collected photogenerated
carriers contribute to a current component that delivers electrical power to
the external circuitry. Therefore, the sweep-out of TPA generated free
carriers can be exploited to generate electrical power.

Sweep-out of carriers and reduction of the effective lifetime has been the
motivation for utilizing a reverse-biased p-n junction
in Raman amplifiers.$^{10-13}$ The forward-bias employed in the present work might seem counterintuitive. However, it should be reminded that reduction of the
effective lifetime can be achieved as long as the diode current is negative.
In other words, a negative voltage is not a prerequisite for carrier
sweep-out. We conducted 2-D numerical drift-diffusion simulations on
structures identical to the fabricated devices using ATLAS (from Silvaco
International), according to a methodology described elsewhere.$^{14}$ It was
verified that the effective lifetime of biases in the fourth quadrant is
about one order of magnitude lower than the open-circuit value. Therefore,
collection of photogenerated carriers in the fourth quadrant of the I-V
characteristics by the built-in field can sufficiently decreases the free
carrier loss and results in CW optical gain.

We now discuss the main limitation of the proposed approach. At very high
optical intensities, the built-in field of the junction is insufficient to
remove the high density of TPA-generated carriers. In such cases, a reverse
bias needs to be applied to increase the field resulting in positive
electrical power dissipation. Therefore, there exists a tradeoff between the
amount of gain (and hence output power) and the electrical power
generation/dissipation.

Finally, high pump intensities, and hence TPA, are also encountered in other
silicon photonic devices that operate based on third-order nonlinear effects
such as Raman lasers$^{9,12}$ as well as in Raman and Kerr-based wavelength
converters.$^{16-18}$ Hence, the present approach is also applicable in these
devices.

In conclusion, two photon absorption and the resulting free carrier
scattering are omnipresent problems in silicon photonic devices that operate
 based on nonlinear optical interactions. It is shown that active removal of
these carriers can be achieved while generating modest amount of electrical
power.

This work is sponsored by DARPA under the EPIC program.

\pagebreak
{\noindent
{\bf References}}

{\noindent
$^1$R. A. Soref, Proceedings of IEEE {\bf 81}, 1687 (1993).}

{\noindent
$^2$L. Pavesi and G. Guillot, {\it Optical interconnects: the Silicon approach},
Springer Series in Optical Sciences, (Springer, 2006; ISBN: 3540289100).}

{\noindent
$^3$D. J. Frank, IBM J. Research and Development {\bf 46}, 235 (2002).}

{\noindent
$^4$International Technology Roadmap for Semiconductors, 2005 Edition,

{\noindent
http://www.itrs.net/.}}

{\noindent
$^5$R. Claps,D. Dimitropoulos, Y. Han, and B. Jalali, Opt. Express {\bf 10},
1305 (2002).}

{\noindent
$^6$R. Claps,D. Dimitropoulos,V. Raghunathan,Y. Han, and B. Jalali,
Opt. Express {\bf 11}, 1731 (2003).}

{\noindent
$^7$R. L. Epinola, J. I. Dadap, R. M. Osgood, Jr., S. J. McNab, and Y. Vlasov,
Opt. Express {\bf 12}, 3713 (2004).}

{\noindent
$^8$X. Qianfan, V. R. Almeida, and M. Lipson, Opt. Express {\bf 12}, 4437 (2004).}

{\noindent
$^9$O. Ozdal and B. Jalali, Opt. Express {\bf 12}, 5269 (2004).}

{\noindent
$^{10}$T. K. Liang, H. K. Tsang, Appl. Phys. Lett. {\bf 84}, 2745 (2004).}

{\noindent
$^{11}$A. Liu, H. Rong, M. Paniccia, O. Cohen, and D. Hak, Opt. Express {\bf 12}, 4261
(2004).}

{\noindent
$^{12}$H. Rong, R. Jones, A. Liu, O. Cohen, D. Hak, A. Fang and M. Paniccia,
Nature {\bf 433}, 725 (2005).}

{\noindent
$^{13}$R. Claps, V. Raghunathan, D. Dimitropoulos, and B. Jalali, Opt. Express {\bf 12}, 2774 (2004).}

{\noindent
$^{14}$D. Dimitropoulos, S. Fathpour, and B. Jalali, Appl. Phys. Lett. {\bf 87}, 261108 (2005).}

{\noindent
$^{15}$S. Fathpour, O. Boyraz, D. Dimitropoulos, and B. Jalali,
IEEE Conf. on Lasers and Electro-optics, CLEO 2006, Long Beach, CA (2006).}

{\noindent
$^{16}$R. Claps, V. Raghunathan, D. Dimitropoulos, and B. Jalali, Opt. Express 11, 2862 (2003).}

{\noindent
$^{17}$R. Espinola, J. Dadap, R. Osgood, Jr., S. McNab, and Y. Vlasov, Opt.
Express {\bf 13}, 4341 (2005).}

{\noindent
$^{18}$H. Fukuda, K. Yamada, T. Shoji, M. Takahashi, Tai Tsuchizawa, T. Watanabe, J. Takahashi, and S. Itabashi, Opt. Express {\bf 13}, 4629-37 (2005).}
\pagebreak

{\noindent
{\bf Figure Captions}}

{\noindent
FIG. 1. Schematic of the fabricated silicon-on-insulator electrooptic
modulator consisting of a p-n junction straddling the waveguide with H=2.0
and h=0.9 $\mu$m. Values of d and W are described in the text.}

{\noindent
FIG. 2. Measured on-off Raman gain versus coupled pump power in the devices
of Fig. 1 at different biasing conditions.}

{\noindent
FIG. 3. Generated electrical power at different coupled pump powers for two
biases in the fourth quadrant.}

{\noindent
FIG. 4. Current-voltage (I-V) characteristics of the diode measured with
a curve-tracer for an input (uncoupled) optical illumination of ~1.1 W.}

\pagebreak

\begin{figure}
\epsfig{file=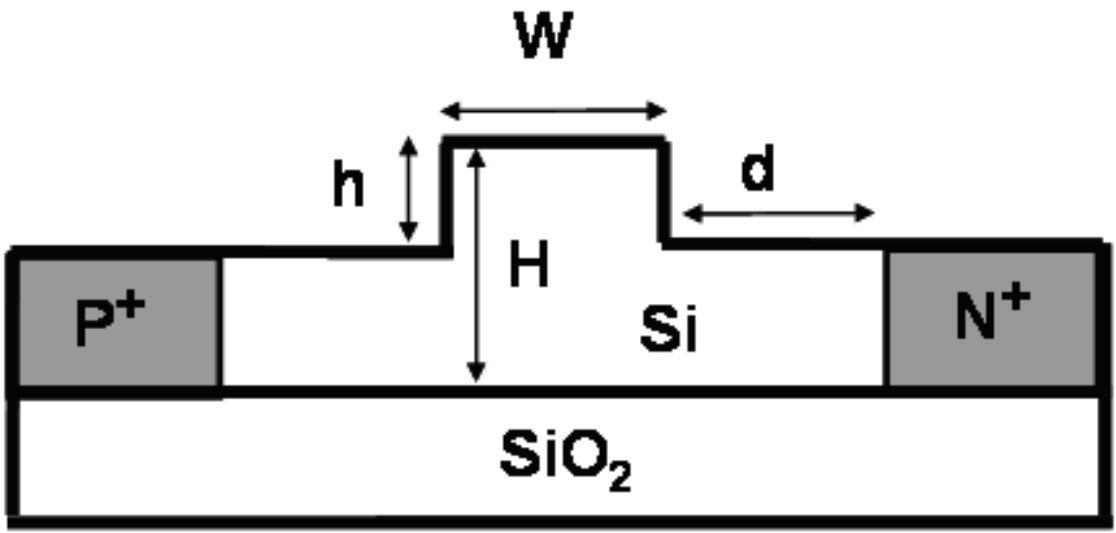, width=15cm }
 \caption{Fathpour
{\it et al.}} \vspace{20cm}
\end{figure}

\begin{figure}

\epsfig{file=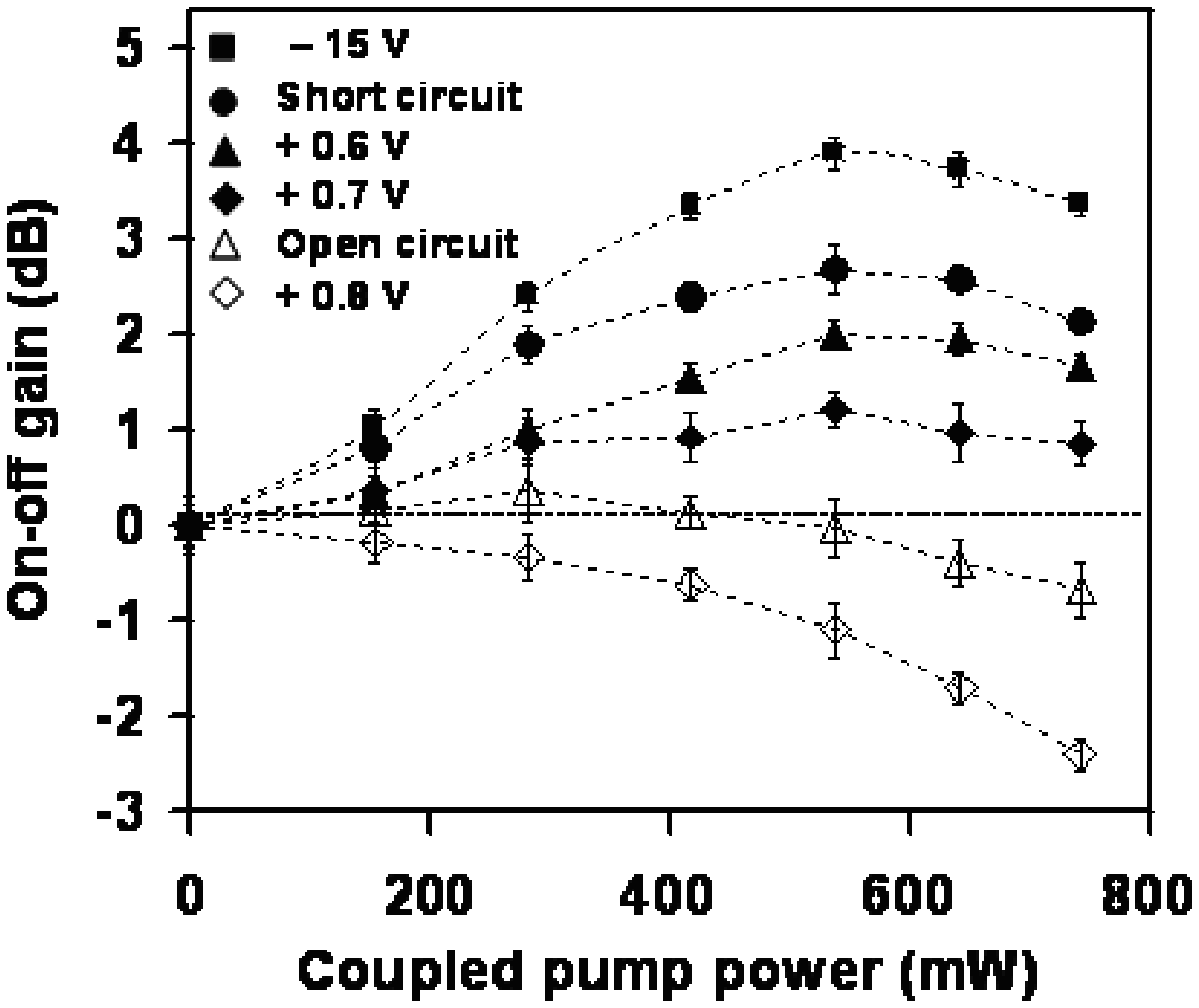, width=15cm } \caption{Fathpour
{\it et al.}}
\end{figure}

\pagebreak

\begin{figure}
\epsfig{file=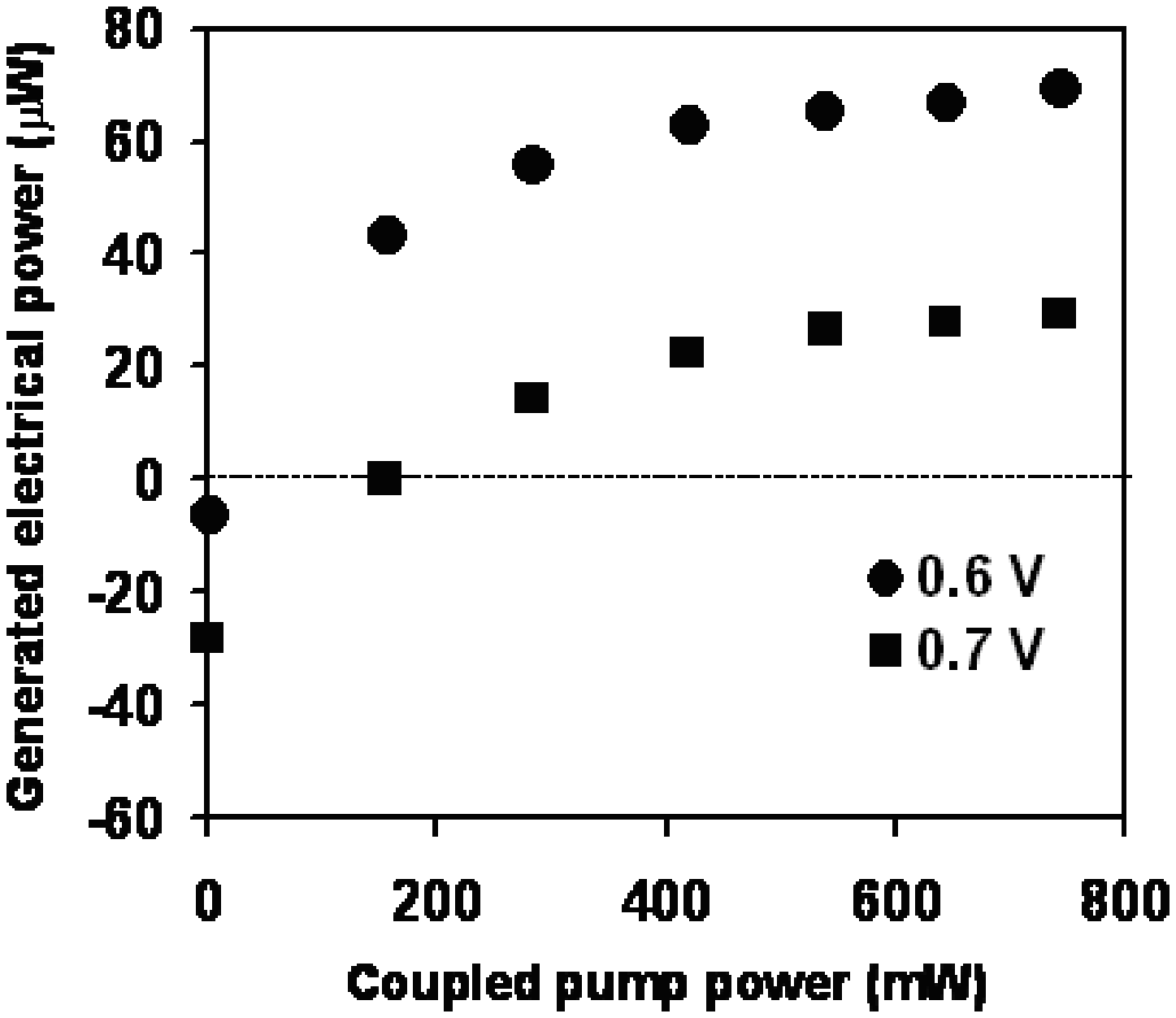, width=15cm }
\caption{Fathpour
{\it et al.}}
\end{figure}

\pagebreak

\begin{figure}
\epsfig{file=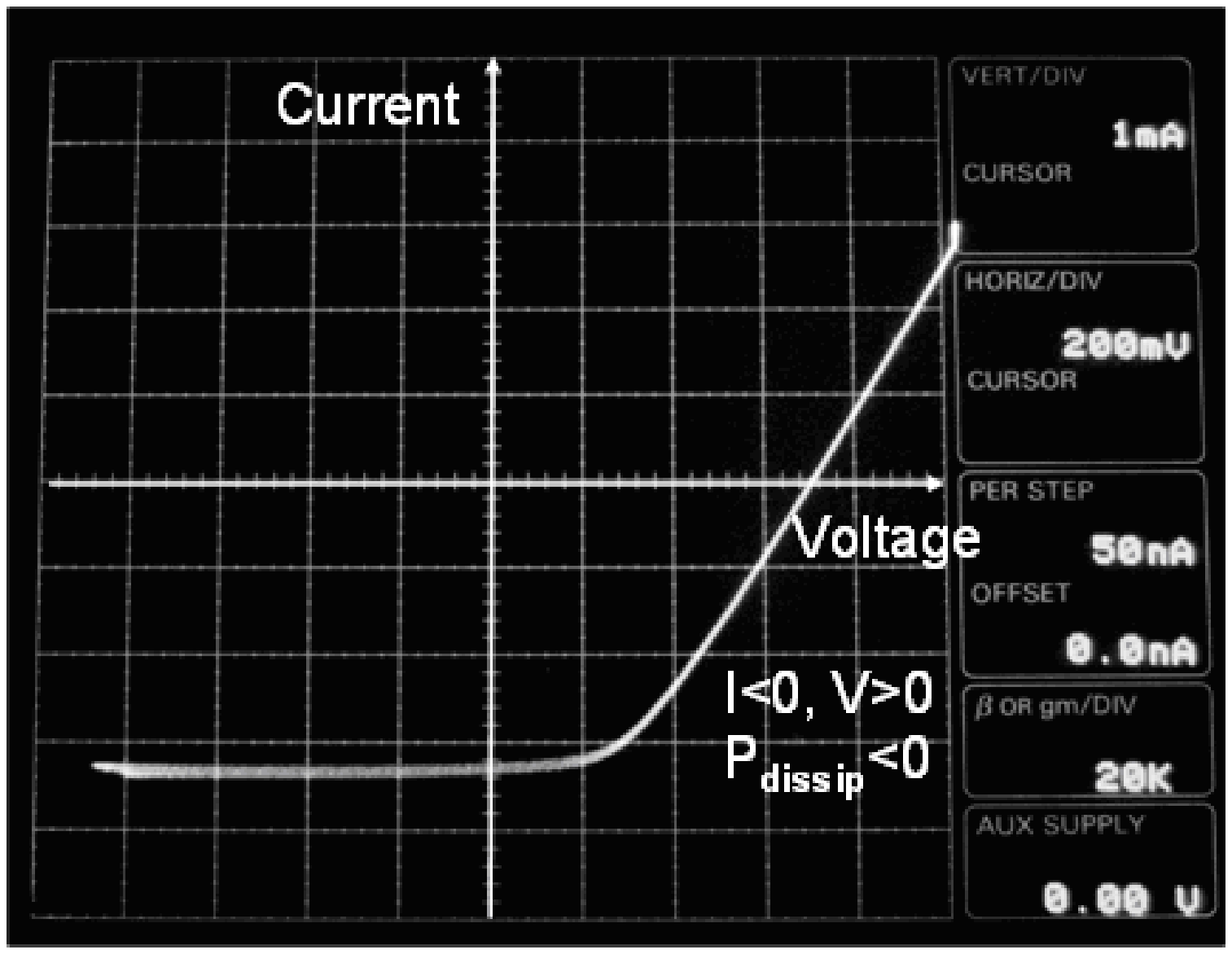, width=15cm }
\caption{Fathpour
{\it et al.}}
\end{figure}

\end{document}